# Electromagnetic $N - \Delta$ transition form factors in a light-front constituent quark model

F. Cardarelli[a], E. Pace[a,b], G. Salmè[c], S. Simula[c]

[a]Istituto Nazionale di Fisica Nucleare, Sezione Tor Vergata
Via della Ricerca Scientifica 1, I-00133 Roma, Italy
[b] Dipartimento di Fisica, Università di Roma "Tor Vergata"
Via della Ricerca Scientifica 1, I-00133 Roma, Italy
[c]Istituto Nazionale di Fisica Nucleare, Sezione Sanità,
Viale Regina Elena 299, I-00161 Roma, Italy

**Abstract**
A parameter-free evaluation of the $N - \Delta$ electromagnetic transition form factors is carried out in a light-front constituent quark model, based on $N$ and $\Delta$ eigenstates of a mass operator which reproduces a large set of hadron energy levels. A one-body electromagnetic current, including the constituent quark form factors previously determined within the same approach from nucleon and pion experimental data, is used. Our prediction for the magnetic transition form factor is compared with the results both of phenomenological analyses of experimental data for the $\Delta$ electroproduction, and of an analysis which includes the effects of off-energy-shell $\pi - N$ final state interaction.

The electromagnetic (e.m.) excitations of nucleon resonances, and in particular the transition to the lowest excited state, the relatively isolated $\Delta$ resonance, can provide useful information on the mechanism responsible for confining quarks and gluons into hadrons, and on the transition to the perturbative regime of QCD [1]. The difficulties to be faced in the non-perturbative regime have motivated the development of effective theories and models, such as constituent quark (CQ) models. Thus, in the last few years the e.m. $N - \Delta$ transition has been studied within non-relativistic or relativized CQ models, including also the effects of configuration mixing [2], and within light-front CQ models, using a simple ansatz for the wave functions [3, 4, 5]. These analyses have shown that relativistic effects play a relevant role, even at low values of the momentum transfer. Since in the near future the $\Delta$-electroproduction processes from nucleons and nuclei will be extensively studied at CEBAF [6], it becomes essential to obtain theoretical predictions for these processes in a proper relativistic framework using a realistic description of CQ dynamics.

Aim of this letter is to compute the e.m. $N - \Delta$ transition form factors in a relativistic CQ model within the framework of the light-front hamiltonian dynamics [7], using $N$ and $\Delta$ eigenstates of the same light-front mass operator and a one-body approximation for the e.m. current. To this end, a light-front CQ model, developed in a series of recent papers [8, 9], will be extended to the $\Delta$. The model describes hadrons as systems of constituent quarks, the other degrees of freedom being frozen; in particular for the baryons the CQ's interact with the $q - q$ potential of Capstick and Isgur (CI) [10], which reproduces a large set of baryon energy levels. In our model an effective one-body e.m. current, including Dirac and Pauli form factors for the CQ's, is used (cf. also Ref. [11]). The CQ form factors have been determined in Ref. [9] using as constraints the experimental e.m. form factors of $\pi$, proton and neutron, obtaining a remarkable, coherent description of the experimental data. In this letter, we will present our parameter-free results for the e.m. $N - \Delta$ transition form factors, using the same CQ form factors of Ref. [9], and compare our $N - \Delta$ magnetic transition form factor with the results of analyses of experimental $\Delta$-electroproduction cross sections. Furthermore, we will study the sensitivity of the transition form factors to different $N$ and $\Delta$ wave functions.

In the light-front hamiltonian dynamics (cf. [7]) intrinsic momenta of the CQ's, $k_i$, can be obtained from the momenta $p_i$ in a general reference frame, through the light-front boost $L_f^{-1}(P)$, which transforms the total momentum of the system $P = \sum_{i=1}^{3} p_i$ to the rest frame one, i.e. $L_f^{-1}(P) \, P = (M, 0, 0, 0)$, and gives no Wigner rotation. Namely, one has $k_i = L_f^{-1}(P) \, p_i$ and, obviously, $\sum_{i=1}^{3} \vec{k}_i = 0$. In this formalism a baryon state in the $u - d$ sector, $|\Psi^{T \; T_3}_{J \; J_n \; \pi}, \tilde{P}\rangle$, is an eigenstate of: i) isospin, $T$ and $T_3$; ii) parity, $\pi$; iii) non-interacting, light-front, angular momentum operators $j^2$ and $j_n$, where the vector $\hat{n} = (0, 0, 1)$ defines the spin quantization axis; iv) total light-front baryon momentum $\tilde{P} \equiv (P^+, \vec{P}_\perp) = \tilde{p}_1 + \tilde{p}_2 + \tilde{p}_3$, where $P^+ = P^0 + \hat{n} \cdot \vec{P}$ and the subscript $\perp$ indicates the projection perpendicular to the spin quantization axis; and v) an intrinsic light-front mass operator, $\mathcal{M} = M_0 + \mathcal{V}$. The state $|\Psi^{T \; T_3}_{J \; J_n \; \pi}, \tilde{P}\rangle$ factorizes into $|\Psi^{T \; T_3}_{J \; J_n \; \pi}\rangle \, |\tilde{P}\rangle$ and the intrinsic part $|\Psi^{T \; T_3}_{J \; J_n \; \pi}\rangle$ is eigenstate of the intrinsic mass operator

$$(M_0 + \mathcal{V}) \, |\Psi^{T \ T_3}_{J \ J_n \ \pi}\rangle \; = \; M |\Psi^{T \ T_3}_{J \ J_n \ \pi}\rangle \tag{1}$$

The operator $M_0 = \sum_{i=1}^{3} \sqrt{m_i^2 + \vec{k}_i^2}$ is the free mass operator, $m_i$ the CQ mass, $\mathcal{V}$ a Poincaré invariant interaction, $M$ the baryon mass, and $J(J+1)$, $J_n$ are the eigenvalues of the operators $j^2$, $j_n$, respectively. The intrinsic, light-front angular momentum eigenstate $|\Psi^{T \ T_3}_{J \ J_n \ \pi}\rangle$ can be obtained by means of the unitary operator $\mathcal{R}^{\dagger} = \prod_{j=1}^{3} R^{\dagger}_{Mel}(\vec{k}_j, m_j)$ from the eigenstate $|\psi^{T \ T_3}_{J \ J_n \ \pi}\rangle$ of the *canonical* angular momentum, i.e. $|\Psi^{T \ T_3}_{J \ J_n \ \pi}\rangle = \mathcal{R}^{\dagger} \, |\psi^{T \ T_3}_{J \ J_n \ \pi}\rangle$, with $R_{Mel}(\vec{k}_j, m_j)$ being the generalized Melosh rotation [12]. Since $M_0$ commutes with $\mathcal{R}^{\dagger}$, applying the Melosh rotations to Eq. (1), the mass equation for the states $|\psi^{T \ T_3}_{J \ J_n \ \pi}\rangle$ can be immediately written

$$(M_0 + V) \, |\psi^{T \ T_3}_{J \ J_n \ \pi}\rangle \; = \; M |\psi^{T \ T_3}_{J \ J_n \ \pi}\rangle \tag{2}$$

where the interaction $V = \mathcal{R}\mathcal{V}\mathcal{R}^{\dagger}$ has to be independent of the total momentum $P$ and invariant upon spatial rotations and translations (cf. [7]).

We can identify Eq. (2) with the baryon wave equation proposed by Capstick and Isgur [10]. The interaction $V = \sum_{i \neq j=1}^{3} V_{ij}$ is a sum over the CQ pairs of the CI effective potential $V_{ij}$, composed by a OGE term (dominant at short separations) and a linear confining term (dominant at large separations). It should be pointed out that the mass operator (2) generates a huge amount of high momentum components and also SU(6) breaking terms in the baryon wave function, due to the presence of the OGE part of the interaction [8, 9], whereas in the current literature only the effects of the confinement scale have been considered, through gaussian or power-law wave functions [3, 4, 5, 13, 14]. As in Ref. [10], the values $m_u = m_d = 0.220 \, GeV$ have been adopted.

Disregarding the color degree of freedom, the state $|\psi^{T \ T_3}_{J \ J_n \ \pi}\rangle$ for a baryon in the $u-d$ sector is symmetric with respect to orbital, spin and isospin coordinates and can be written as: $|\psi^{T \ T_3}_{J \ J_n \ \pi}\rangle = \sum_{LS,[f]_\lambda} \sum_{M_L M_S} \langle LM_L, SM_S | JJ_n \rangle \, |\psi^{S[f]_\lambda}_{LM_L}(\vec{k}_1, \vec{k}_2, \vec{k}_3)\rangle \, |SM_S, TT_3, [f]_\lambda\rangle$, where $L$ and $S$ are the total orbital angular momentum and spin, respectively, $[f]_\lambda = [a], [s], [m]_\lambda$ denotes the irreducible antisymmetric, symmetric and mixed-symmetry representations of the $\{\mathcal{S}_3\}$ group, respectively, and $\lambda$ is the usual Yamanouchi symbol.

The mass equation (2) has been solved by expanding the orbital states $|\psi^{S[f]_\lambda}_{LM_L}\rangle$ onto a (truncated) set of harmonic oscillator (HO) basis states in the Jacobi coordinates $\vec{p} = \vec{k}_1$ and $\vec{k} = (\vec{k}_2 - \vec{k}_3)/2$ [5] for the three-quark system (details will be given elsewhere [15]) and applying to the Hamiltonian $M_0 + V$ the Rayleigh-Ritz variational principle. We have checked that the convergence for all the quantities considered in this work can be reached including in the expansion all the basis states up to 20 HO quanta. For the nucleon, the orbital $S$, $S'$ and $D$ components were considered, and the corresponding probabilities are: $P_S^N = 0.981$, $P_{S'}^N = 0.017$, and $P_D^N = 0.002$ (the antisymmetric $L = 0$ component was found to be completely negligible) [9, 15]. For the $\Delta$, the $S$ and $D$ components have been considered and the probabilities are: $P_S^\Delta = 0.989$ and $P_D^\Delta = 0.011$ (an $S'$ component is impossible for the $\Delta$). The $P$ component has been disregarded, since it is weakly coupled by the interaction with the dominant component of the wave function; indeed it was found quite small in Ref. [10]. The calculated masses are $M_N = 0.940 \ GeV$ and $M_\Delta = 1.250 \ GeV$.

In order to investigate the sensitivity of the $N$ and $\Delta$ wave functions on the various components of the interaction, we have solved Eq. (2) by retaining only the linear confining part of the CI interaction, $V_{(conf)}$, or the spin-independent part, $V_{(si)}$, which includes a coulomb-like term, or considering the full interaction, $V_{(CI)}$. The corresponding baryon states will be indicated as $|\psi^{conf}\rangle$, $|\psi^{si}\rangle$, and $|\psi^{CI}\rangle$, respectively. Obviously, only the symmetric orbital $S$ component is present in $|\psi^{conf}\rangle$ and $|\psi^{si}\rangle$, and these states are equal for $N$ and $\Delta$. Let us now consider the CQ momentum distribution defined as $n(p) \equiv \sum_{LS,[f]_\lambda} \int |\psi_{LM_L}^{S[f]_\lambda}(\vec{p}, \vec{k}_2, \vec{k}_3)|^2 \ \delta(\vec{p} + \vec{k}_2 + \vec{k}_3) \ d\Omega_{\vec{p}} \ d\vec{k}_2 \ d\vec{k}_3$. The momentum distribution $n(p)$, times $p^2$, for the full CI interaction is shown in Fig. 1, both for $N$ and $\Delta$, together with the momentum distributions of the corresponding $D$ components. The $D$ component of the nucleon wave function gives a negligible contribution to the momentum distribution in the whole range of momenta shown in Fig. 1, and the $D$ component of the $\Delta$ becomes relevant only for $p > 2 \ GeV/c$. Therefore, in this letter we will neglect the contribution of the $D$ components in the calculation of the $N - \Delta$ transition form factors. In Fig. 1 the momentum distributions $n^{conf}(p)$ and $n^{si}(p)$, corresponding to the states $|\psi^{conf}\rangle$ and $|\psi^{si}\rangle$, respectively, and the momentum distribution of the wave function of Ref. [5], taken as an example of the gaussian wave functions usually employed in light-front calculations, are also reported. As already observed in the case of pseudoscalar and vector mesons [8], the gaussian momentum distribution is very similar to $n^{conf}(p)$, while the coulomb-like spin-independent part of the interaction produces a large amount of high momentum components in the wave function. The spin-dependent OGE part of the $q-q$ interaction, determining the hyperfine splitting of hadron mass spectra (in particular, the $N - \Delta$ mass splitting) generates also a considerable splitting of $n(p)$ in $N$ and $\Delta$ at $p > 1 \ GeV/c$.

In order to compute the $N - \Delta$ transition form factors we will use, as in Ref. [9], the eigenvectors of Eq. (1) and the one-body component of the e.m. current, which for a three quark system is given by

$$\mathcal{I}^\nu(0) = \sum_{j=1}^{3} I_j^\nu(0) = \sum_{j=1}^{3} \left( e_j \gamma^\nu f_1^j(Q^2) \ + \ i\kappa_j \frac{\sigma_\mu^\nu q^\mu}{2m_j} f_2^j(Q^2) \right) \qquad (3)$$

where $\sigma_\mu^\nu = \frac{i}{2}[\gamma^\nu, \gamma_\mu]$, $e_j$ is the charge of the j-th quark, $\kappa_j$ the corresponding anomalous magnetic moment, $f_{1(2)}^j$ its Dirac (Pauli) form factor, and $Q^2 \equiv -q \cdot q$ the squared four-momentum transfer. It is well known that the hadron e.m. current has to include two-body components for fulfilling gauge and rotational invariances (see Ref. [7]). However, in Ref. [9] we have shown that the effective one-body current (Eq. (3)) is able to give a coherent description of both the $\pi$ and $N$ experimental form factors. Therefore in this letter the one-body component of the e.m. current with the same CQ form factors, $f_{1(2)}^u$ and $f_{1(2)}^d$, and the same anomalous magnetic moments, $\kappa_u$ and $\kappa_d$, of Ref. [9] will be adopted.

The e.m. $N - \Delta$ transition current matrix elements can be written as follows [16, 3]

$$\langle \Psi^{\frac{3}{2}\tau_\Delta}_{\frac{3}{2}\nu_\Delta +1}, \tilde{P}_\Delta | \mathcal{I}^\nu(0) | \Psi^{\frac{1}{2}\tau_N}_{\frac{1}{2}\nu_N +1}, \tilde{P}_N \rangle = \langle \frac{1}{2}\tau_N, 10 | \frac{3}{2}\tau_\Delta \rangle \; \bar{w}^\mu(\tilde{P}_\Delta, \nu_\Delta) \; \Gamma_\mu^{\;\nu} \; u(\tilde{P}_N, \nu_N) \qquad (4)$$

where $\bar{w}^\mu(\tilde{P}_\Delta, \nu_\Delta)$ is the Rarita-Schwinger spin-$\frac{3}{2}$ spinor, $u(\tilde{P}_N, \nu_N)$ the nucleon spinor and the Clebsch-Gordan coefficient $\langle \frac{1}{2}\tau_N, 10 | \frac{3}{2}\tau_\Delta \rangle = \sqrt{2/3} \; \delta_{\tau_\Delta \tau_N}$ defines explicitely the isospin coupling [17]. The tensor $\Gamma_{\mu\nu}$ can be cast as a linear combination of three independent, covariant tensors $G^M_{\mu\nu}$, $G^E_{\mu\nu}$, and $G^C_{\mu\nu}$ multiplied by the magnetic, $G_M$, electric, $G_E$, and Coulomb, $G_C$, transition form factors: $\Gamma_{\mu\nu} = G_M(Q^2) \; G^M_{\mu\nu} + G_E(Q^2) \; G^E_{\mu\nu} + G_C(Q^2) \; G^C_{\mu\nu}$ (see, e.g., Refs. [16, 3]). In the light-front formalism (see, e.g., Ref. [14]) the spacelike e.m. form factors are related to the *plus* component of the e.m. current, $\mathcal{I}^+ = \mathcal{I}^0 + \hat{n} \cdot \vec{\mathcal{I}}$, evaluated in a frame where $q^+ = q^0 + \hat{n} \cdot \vec{q} = P_\Delta^+ - P_N^+ = 0$; such a choice allows to suppress the contribution of the pair creation from the vacuum [18]. It is important to realize that within the light-front framework, since the e.m. form factors can be calculated using the matrix elements of $\mathcal{I}^+$ only, our one-body approximation refers to the $\mathcal{I}^+$ component of the current; this means that, with a suitable definition of the other components, the e.m. current can fulfil current conservation [5, 19].

The transition form factors can be obtained by considering three independent matrix elements of $\mathcal{I}^+$ and their linear relations with $G_M$, $G_E$, and $G_C$. Following Ref. [3] the matrix elements $\mathcal{I}_{\frac{3}{2},\frac{1}{2}}$, $\mathcal{I}_{\frac{1}{2},\frac{1}{2}}$, and $\mathcal{I}_{\frac{1}{2},-\frac{1}{2}}$ will be used, with the short-hand notation $\langle \Psi^{\frac{3}{2}\tau}_{\frac{3}{2}\nu_\Delta +1}, \tilde{P}_\Delta | \mathcal{I}^+(0) | \Psi^{\frac{1}{2}\tau}_{\frac{1}{2}\nu_N +1}, \tilde{P}_N \rangle = \mathcal{I}_{\nu_\Delta, \nu_N}$. The expressions of the transition form factors are the following:

$$G_M(Q^2) = B \; \{ \mathcal{I}_{\frac{3}{2},\frac{1}{2}} \sqrt{3} \; [Q^4 + 2Q^2 M_N^2 + m_+^2 m_-^2] +$$
$$+ \mathcal{I}_{\frac{1}{2},\frac{1}{2}} \; q_L \; [Q^2(3M_\Delta - M_N) - m_+ m_-^2] + \mathcal{I}_{\frac{1}{2},-\frac{1}{2}} \; [Q^2(2M_\Delta^2 + m_-^2) + m_+^2 m_-^2] \} \quad (5)$$

$$G_E(Q^2) = B \; \{ \mathcal{I}_{\frac{3}{2},\frac{1}{2}} \; 3^{-1/2} \; [Q^4 - 2Q^2(2M_\Delta^2 - M_N^2) + m_+^2 m_-^2] +$$
$$+ \mathcal{I}_{\frac{1}{2},\frac{1}{2}} \; q_L \; m_+ \; [Q^2 - m_-(3M_\Delta + M_N)] + \mathcal{I}_{\frac{1}{2},-\frac{1}{2}} \; [Q^2(m_+ m_- + 2M_\Delta M_N) - m_+^2 m_-^2] \} \quad (6)$$

$$G_C(Q^2) = B \; 2M_\Delta \; \{ \mathcal{I}_{\frac{3}{2},\frac{1}{2}} \sqrt{3} \; M_\Delta \; [-Q^2 + m_+ m_-] +$$
$$+ \mathcal{I}_{\frac{1}{2},\frac{1}{2}} \; q_L \; [-m_+^2 - 2m_+ m_- + Q^{-2} \; m_+^2 m_-^2] + \mathcal{I}_{\frac{1}{2},-\frac{1}{2}} \; [Q^2 M_N - (2M_\Delta + M_N) \; m_+ m_-] \} \quad (7)$$

where $B = M_N \; [(Q^2 M_N + m_+ m_-^2) \; m_+ \; q_L]^{-1}$, $q_L = q_x - iq_y$ and $m_\pm = M_\Delta \pm M_N$. A detailed analysis of our results, obtained using the eigenstates of the mass operator of Eq. (1) and including the $D$ components, as well as of the violation of the so-called "angular condition" [5], will be presented in a forthcoming paper, where the explicit expression for the $N - \Delta$ transition current matrix elements will also be given [20]. We point out that the violation of the angular condition within our approach has been found as small as for simple gaussian wave functions and point-like CQ's [5]. The numerical calculations of the matrix elements $\mathcal{I}_{\nu_\Delta, \nu_N}$ involve multifold integrations, carried out through a Monte Carlo routine [21].

The results obtained for $G_M(Q^2)$ (Eq. (5)), divided by the dipole form factor $3G_D(Q^2) = 3(1 + Q^2/0.71)^{-2}$, assuming point-like CQ's, are reported in Fig. 2. The solid, short-dashed and dotted lines correspond to solutions of Eq. (1) using $V_{(CI)}$, $V_{(si)}$ and $V_{(conf)}$, respectively. For a comparison the result obtained from Eq. (5) with the gaussian wave function of Ref. [5] is reported. The figure shows a huge effect due to the configuration mixing in the baryon wave function, leading to a strong deviation from the dipole behaviour at variance with existing calculations [2, 3, 4, 5]. In general the larger is the high momentum content of the $N$ and $\Delta$ wave functions, the larger the magnetic transition form factor at high $Q^2$. It has to be noticed, however, that the $S'$ component of the nucleon wave function, which dominates the CQ momentum distribution at $p > 2\ GeV/c$ in the nucleon [15], produces a small but non negligible reduction of $G_M(Q^2)$ (compare the solid and long-dashed lines in Fig. 2). The small differences between $G_M(Q^2)$ calculated using $|\psi^{si}\rangle$ and $|\psi^{CI}\rangle$ can be easily explained, since the matrix elements (4) involve the product of $N$ and $\Delta$ wave functions, which is very similar when the $|\psi^{si}\rangle$ and $|\psi^{CI}\rangle$ states are considered.

The $N-\Delta$ transition form factor $G_M$ obtained using the eigenstates of Eq. (1) with the CI interaction and the one-body component of the e.m. current (Eq. (3)) with the CQ form factors of Ref. [9], is shown in Fig. 3. One can immediately note that the CQ form factors largely reduce $G_M(Q^2)$ at high $Q^2$ and that a faster than dipole fall-off is found. In Fig. 3 the results of phenomenological analyses of the experimental data for the $\Delta$ electroproduction up to $Q^2 = 11\ (GeV/c)^2$ are reported, showing a fall-off faster than the predictions of the quark counting rules [1, 22, 23]. In such analyses the magnetic transition form factor is extracted from the experimental (e, e') cross sections by considering an incoherent sum of Breit-Wigner terms, which yield the resonant contributions, and a phenomenological background, without any dynamical model for the $\pi - N$ final-state interaction. As an example of a microscopical analysis, the result obtained in Ref. [17] from the data at low values of $Q^2$ is also reported in Fig. 3. In Ref. [17] the inclusive and exclusive $\pi$ electroproduction in the $\Delta$-resonance region is described by the coherent sum of the $\Delta$-excitation term and of the Born contributions, evaluated within a dynamical hamiltonian model, including in the final state the $\pi - N$ off-energy-shell interaction. The parameters of the model are fixed from previous works, while $G_M(Q^2)$ is determined from a fit to the experimental data. The results of Ref. [17] are substantially lower than the corresponding results obtained in Refs. [1, 22, 23] at low $Q^2$; such a difference might be due to a resonant contribution from the Born terms, generated by the $\pi - N$ final state interaction. Our parameter-free result for $G_M(Q^2)$ differs at most 20% from the results of both the phenomenological and the microscopical analyses at low values of $Q^2$. At high values of the momentum transfer, where the $\pi - N$ final-state interaction could be of minor relevance, our result is close to the phenomenological ones.

Finally, we report in Fig. 4 the ratio $E_2/M_1 = -G_E/G_M$ (Eqs. (5,6)) and the coulomb transition form factor $G_C$ (Eq. (7)). Since both $D$ components and two-body e.m. currents are not included in the present calculation of these small quantities, our results are intended to have only an exploratory nature. The solid and dashed lines correspond to the eigenstates of Eq. (1) for $N$ and $\Delta$ with CI interaction and to the e.m. current of Eq. (3), including the CQ form factors of Ref. [9] or assuming point-like constituent quarks, respectively. The non-zero values of $E_2/M_1$ and $G_C$, evaluated using only the $L = 0$ component of the wave functions, represent the effects of the relativistic nature of our calculation. From Fig. 4 it is clear that the CQ form factors do not change at all the ratio $E_2/M_1$, although they have a large effect on $G_M$, $G_E$ and $G_C$ separately (compare for $G_M$ the solid lines in Figs. 2 and 3). The quite different result obtained for $E_2/M_1$ from Eqs. (5,6) with the gaussian wave function of Ref. [5] and point-like CQ's, is also reported in Fig. 4 and shows that this ratio could provide information on the high-momentum components of the $N$ and $\Delta$ states, independently of the possible e.m. structure of the CQ's.

Summarizing, we have evaluated with no free parameter the $N - \Delta$ transition form factors within the framework of the light-front hamiltonian dynamics. We have used the $N$ and $\Delta$ eigenstates of the same mass operator and a one-body approximation for the e.m. current, containing the CQ form factors already determined in Ref. [9], where the pion and nucleon experimental form factors are nicely described within the same approach. The calculated magnetic transition form factor falls between the phenomenological and the microscopical analyses at low values of $Q^2$ and follows the faster than dipole decrease of the available experimental data at high values of $Q^2$. In order to draw more definite conclusions, the study of the effects of the small $D$ component of the $\Delta$ wave function, as well as of the possible role of two-body e.m. currents, is in progress.

**Acknowledgments.** We are very indebted to S. Capstick for many useful discussions and we thank C. Keppel for providing us with the results of her analysis of experimental cross sections.

# Figure Captions

Fig. 1. The momentum distribution $n(p)$ of the constituent quarks in $N$ and $\Delta$, times $p^2$. The long-dashed and solid lines are the momentum distributions obtained from the eigenstates of Eq. (2) corresponding to $V_{(CI)}$ [10] for $N$ and $\Delta$, respectively; the long-dashed and solid lines marked by open dots are the momentum distributions of the corresponding $D$ components. The dotted and dashed lines are the momentum distributions $n^{conf}(p)$ and $n^{si}(p)$ corresponding to the eigenstates of Eq. (2) obtained with $V_{(conf)}$ or $V_{(si)}$, respectively. The dot-dashed line is the momentum distribution calculated with the gaussian wave function of Ref. [5].

Fig. 2. The magnetic $N - \Delta$ transition form factor $G_M(Q^2)/3G_D(Q^2)$ vs. $Q^2$, with $G_D(Q^2) = (1+Q^2/0.71)^{-2}$. $G_M(Q^2)$ is obtained from Eq. (5) using the eigenstates of Eq. (1) for $N$ and $\Delta$, and assuming point-like CQ's (i.e. $f_1^{u(d)} = 1$ and $\kappa_{u(d)} = 0$). The solid, dashed and dotted lines correspond to $V_{(CI)}$ [10], $V_{(si)}$ and $V_{(conf)}$, respectively. The long-dashed line is $G_M/3G_D$ obtained by neglecting the $S'$ component of the nucleon eigenstate $|\psi^{CI}\rangle$ of Eq. (2). The dot-dashed line is the result calculated from Eq. (5) with the gaussian wave function of Ref. [5].

Fig. 3. The magnetic $N - \Delta$ transition form factor $G_M(Q^2)/3G_D(Q^2)$ vs. $Q^2$. The solid line is obtained from Eq. (5) using: i) the eigenstates of Eq. (1) corresponding to the full interaction $V_{(CI)}$ [10], both for the $N$ and the $\Delta$; ii) the plus component of the e.m. current of Eq. (3); iii) the CQ form factors of Ref. [9]. The data are taken from the phenomenological analyses of the experimental $\Delta$ electroproduction cross sections of Ref. [22] (open triangles), Ref. [1] (full squares), and Ref. [23] (open dots). To obtain $G_M(Q^2)$ the quantity $G_M^*(Q^2)$ of Ref. [22] has been multiplied by $\sqrt{1 + Q^2/(M_\Delta + M_N)^2}$, and the quantity $G_T(Q^2) = F_\Delta(Q^2)$ of Refs. [1] and [23], disregarding the contribution of $G_E(Q^2)$, has been multiplied by $\sqrt{Q^2/(Q^2 + \nu^2)} \sqrt{1 + Q^2/(M_\Delta + M_N)^2}$, with $\nu = (M_\Delta^2 - M_N^2 + Q^2)/2M_N$. The dotted line is the result for $G_M(Q^2)$ obtained in Ref. [17] from experimental data at low values of $Q^2$ within a dynamical model of inclusive and exclusive $\pi$-electroproduction in the $\Delta$-resonance region.

Fig. 4. (a) The ratio $E_2/M_1 = -G_E/G_M$ vs. $Q^2$, obtained from Eqs. (5,6) using the eigenstates of Eq. (1) for $N$ and $\Delta$ with the CI interaction, and the one-body e.m. current of Eq. (3), either including the CQ form factors of Ref. [9] (solid line) or assuming point-like CQ's (dashed line); the two lines largely overlap. The dot-dashed line is the result obtained from Eqs. (5,6) with the gaussian wave function of Ref. [5] and point-like CQ's. (b) The same as in (a), but for the $N - \Delta$ coulomb transition form factor $G_C(Q^2)$ obtained from Eq. (7).

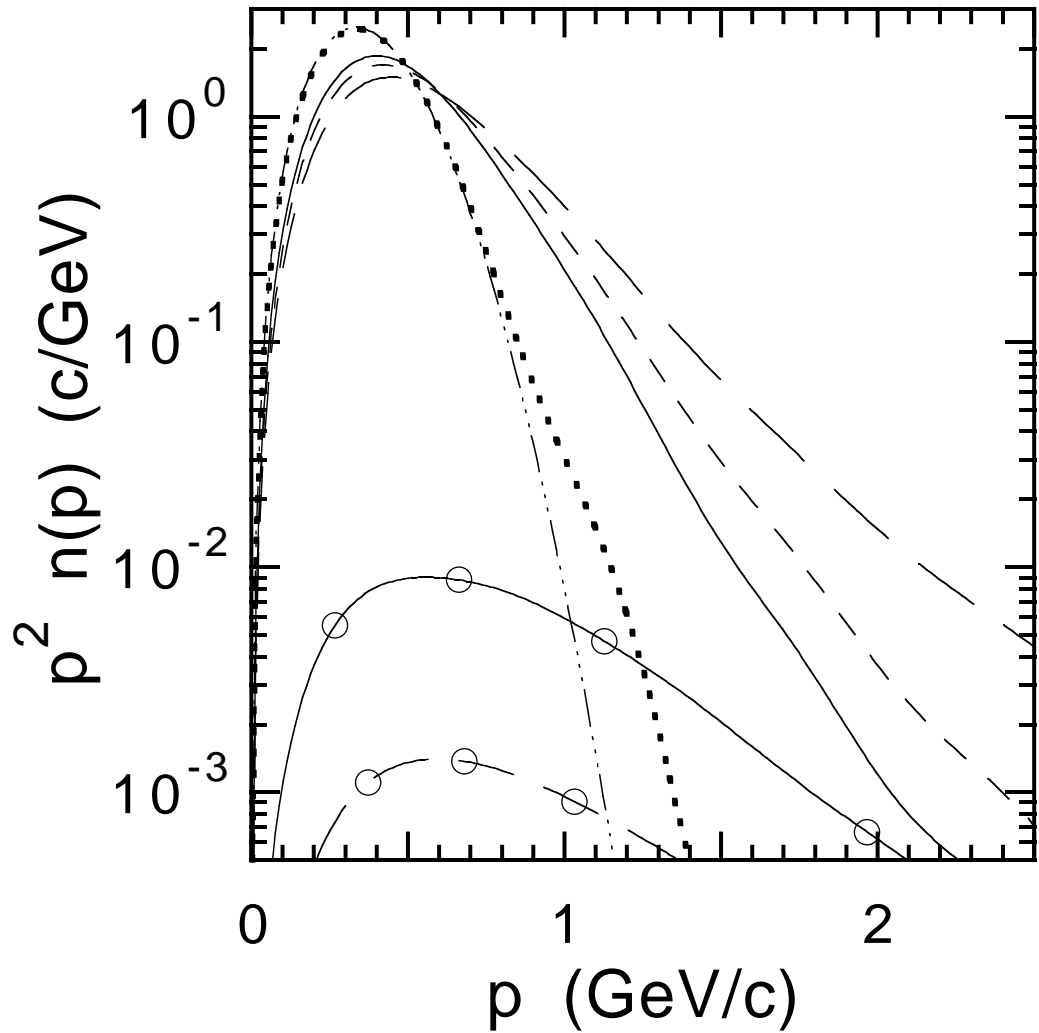

Fig. 1 - F. Cardarelli, E. Pace, G. Salmè and S. Simula. Phys. Lett. B.

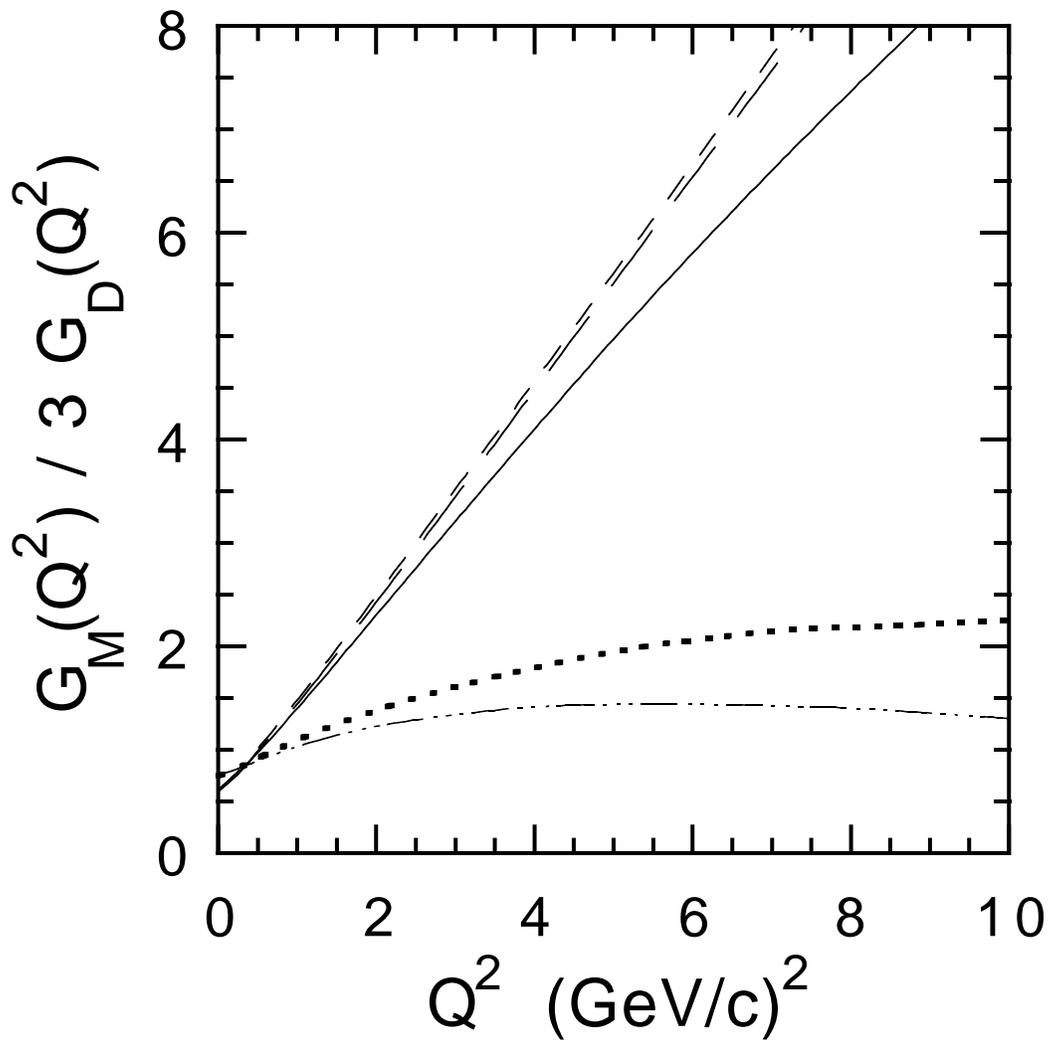

Fig. 2 - F. Cardarelli, E. Pace, G. Salmè and S. Simula. Phys. Lett. B.

Fig. 3 - F. Cardarelli, E. Pace, G. Salmè and S. Simula. Phys. Lett. B.

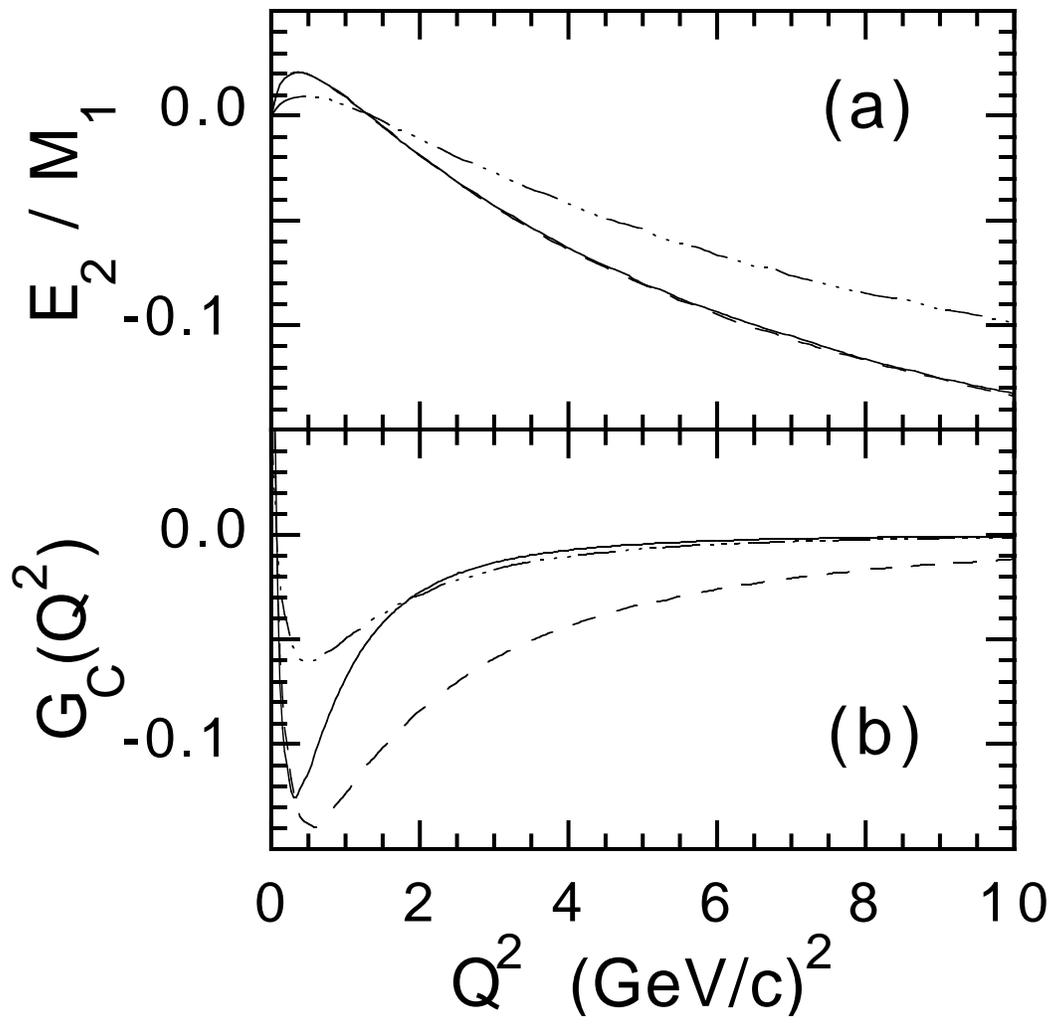

Fig. 4 - F. Cardarelli, E. Pace, G. Salmè and S. Simula. Phys. Lett. B.

13